\documentclass[aps,prl,showpacs,preprintnumbers,%
               twocolumn,groupedaddress,amsfonts]{revtex4}
\usepackage{graphicx}
\usepackage{dcolumn}
\usepackage{bm}
\usepackage{amssymb,amsmath}%
\usepackage{color}
\usepackage{multirow}
\usepackage{amsmath}
\usepackage{epsfig}
\usepackage{subfigure}
\usepackage{float}
\usepackage{verbatim} 
\usepackage{booktabs}
\usepackage[english]{babel}
\usepackage[latin1]{inputenc}
\usepackage{times}
\usepackage[T1]{fontenc}

\suppressfloats[!]

\begin{document}
\preprint{FERMILAB-PUB-11-218-T}
\title{NLO predictions for a lepton, missing transverse momentum and dijets at the Tevatron}

\author{John M. Campbell, Adam Martin and Ciaran Williams \\
}

\address{
Fermilab National Accelerator Laboratory \\
Batavia, Illinois 
\\
USA 60510.}
\begin{abstract}
In this letter we investigate the various processes that can contribute to a final state consisting
of a lepton, missing transverse momentum and two jets at Next to Leading Order (NLO) at the Tevatron. 
In particular we consider the production of $W/Z + 2$ jets, diboson pairs, single top and the
$t\overline{t}$ process with both fully leptonic and semi-leptonic decays. We present  distributions
for the invariant mass of the dijet system and normalisations of the various processes, accurate to
NLO.
\end{abstract}
\date{\today}
\maketitle

\section{Introduction}
\label{sec:Intro}

CDF has recently reported an excess in the dijet invariant mass distribution around $150$~GeV for events containing a lepton, missing energy and exactly two jets~\cite{Aaltonen:2011mk}. 
This has led to many new physics interpretations~\cite{Eichten:2011sh,Buckley:2011vc,Yu:2011cw,Kilic:2011sr,Cheung:2011zt,He:2011ss,Wang:2011ta,Sato:2011ui,Nelson:2011us,
Anchordoqui:2011ag,Dobrescu:2011px,Ko:2011ns,Fox:2011qd,Jung:2011ue,Chang:2011wj,Bhattacherjee:2011yh,Cao:2011yt,Carpenter:2011yj,
Enkhbat:2011qz} as well 
as some suggestions that the excess can  be explained within the framework of  the
Standard Model (SM)~\cite{Plehn:2011nx,Sullivan:2011hu}.
Both of the proposed Standard Model explanations require changes in the normalisation of backgrounds
containing top quarks, which could be the result of higher order perturbative corrections that have
not been fully accounted for in the analysis. In this letter we investigate this issue by performing
a NLO study of SM processes that can produce the final state of interest. We investigate
the cross sections and invariant mass distributions under the CDF cuts and assign an estimate of the
corresponding theoretical uncertainty. To provide a further calibration of these processes, we vary
the cuts by allowing additional jets in the final state and by changing the minimum jet transverse
momentum ($p_T$).

\section{Description of the calculation} 
\label{sec:Descrip}

We present results for NLO cross sections using the latest version (v6.0) of 
MCFM~\cite{Campbell:2011bn}. This includes NLO contributions to the production of
$V + 2$~jets (where $V=W,Z$), $VV$~\footnote{In this letter we neglect  $ZZ(\rightarrow q\overline{q})$ 
since its contribution in the mass range of interest is negligible.}, $t\overline{t}$ and single top. 
The current implementation of MCFM does not include radiation of gluons from quarks produced in the decay of vector bosons
or top quarks.
Such contributions should not lead to additional peaks in the dijet invariant mass distribution, particularly
in the region of interest, and we expect that our 
current implementation provides a satisfactory description of any potential features in these processes.

We perform our calculation using the cuts described in 
ref.~\cite{Aaltonen:2011mk}.  Namely, we ask for events containing 
exactly two jets with $p_{T}^j > 30$~GeV in $|\eta^j| < 2.4$ units of rapidity.
We use the $k_{T}$ algorithm with parameter $R = 0.4$, but have checked that
differences with the cone algorithm are small. 
The jets must be separated by at most $2.5$ units of rapidity and 
the transverse momentum of the dijet system ($p^{jj}_T$) is constrained by $p^{jj}_T > 40$~GeV.
Events should contain exactly one lepton in $|\eta^{\ell}| < 1$, $p^{\ell}_T > 20$~GeV that is
separated from the jets by $R_{j\ell} > 0.52$. We require that the missing transverse momentum ($E_T^{\rm miss}$) satisfies
$E_T^{\rm miss} > 25$~GeV and is separated azimuthally from the leading jet, $\Delta\phi >0.4$.

We evaluate cross sections using the MSTW08 PDF set~\cite{Martin:2009iq} (matched to the appropriate order in
perturbation theory) , using the default set of electroweak parameters in MCFM~\cite{Campbell:2011bn}.
For simplicity we choose to evaluate
all of the processes at a scale of $2m_W$. Such a scale is motivated by its proximity to both the top
mass and the region of the observed excess. In the estimate of the theoretical uncertainty we vary this by
a factor of two in each direction, thus encompassing many typical scale choices for each of the
processes considered.

For the vector boson decays we include two families of leptons ($e$
and $\mu$) in  each process. When there are two leptons present (such as in $Z +$ jets) we require that one
lepton satisfies the rapidity and momentum cuts and the other is missed to provide a source of $E_T^{\rm miss}$.
Jets that do not lie within our acceptance are not included as missing energy.
For the $W(\ell\nu)Z(q\overline{q})$ process we impose an artificial $m_{q\overline{q}} > 10$ GeV cut in 
order to avoid the production of real photons, a contribution that should be included in the QCD background
that we do not attempt to model here.
We have checked predictions for the dijet invariant mass from single top production using both the four and
five flavour schemes at LO and observe no distinctive shape difference between the two. The results presented
in this paper are in the five-flavour scheme, which yields a larger cross section and could therefore be
viewed as the more conservative choice.

\section{Results} 
\label{sec:res} 

\begin{center} 
\begin{figure} 
\includegraphics[width=7cm]{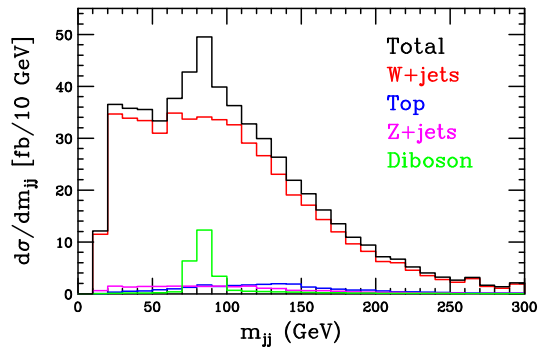}
\caption{NLO predictions for $m_{jj}$ using the CDF cuts (exactly two jets). The combination of $t{\bar t}$
and single top backgrounds is denoted ``Top'', while the contributions from vector boson pairs have been
summed and are indicated by ``Diboson''.} 
\label{fig:mjj30Ex}
\end{figure}
\end{center}

In Table~\ref{CrosssecEx} we present the results for cross sections obtained using the CDF cuts described in
the previous section.  We show results for each contribution separately at LO and NLO and also the
ratio of NLO to LO (the $K$-factor). We find  that under the cuts the NLO corrections to many of the
processes are relatively small, i.e. the $K$-factors are close to unity. The exceptions to this are the
diboson processes, which all receive corrections larger than $50\%$ due to the introduction of events
where one of the hard jets arises from the real radiation.
At NLO the estimated theoretical uncertainty is $11$\% or less for all processes.
The invariant mass of the dijet system is depicted in Fig.~\ref{fig:mjj30Ex}, where we observe that above the $W$ and $Z$
resonance region the $W+2$~jets process is a falling distribution while the top contribution peaks
around 140~GeV. The $Z+2$ jets process has the same shape as the $W+2$ jets background. 

\begin{center} 
\begin{figure} 
\includegraphics[width=7cm]{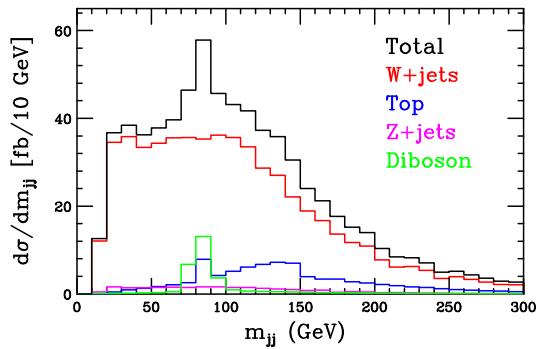}
\caption{NLO predictions for $m_{jj}$ using the ``inclusive'' CDF cuts (two or more jets). The labelling
is as in Fig.~\ref{fig:mjj30Ex}.} 
\label{fig:mjj30Inc}
\end{figure}
\end{center}

Applying the two-jet exclusivity requirement in our fixed order calculation leaves the predictions susceptible
to potentially-large logarithms and significant higher-order corrections.
For this reason, in Table~\ref{Crosssec30Inc} we also present results for cross sections at NLO using the same cuts as
before, but without the two-jet exclusivity requirement (``inclusive'').  For these results the jet cuts described earlier
apply only to the two hardest jets. At LO this only affects
$t\overline{t}$ production with semi-leptonic decays since this process contains four partons.
Dropping the 2-jet requirement increases the cross section by around a factor of seven. For all processes,
additional radiation present in the NLO calculations is no longer cut away, resulting in larger $K$-factors.
The estimated uncertainty from scale variation is very similar to the exclusive case.
The invariant mass of the two hardest jets is presented in Fig.~\ref{fig:mjj30Inc}, where the increase in the top
background is clear. Moreover, there is a clear kinematic feature in the top backgrounds in the region of $150$~GeV.
This edge arises from the semi-leptonic decay of a $t\overline{t}$ pair, as pointed out in
ref.~\cite{Plehn:2011nx} in the context of a source for the CDF excess.
In the total $m_{jj}$ distribution this edge manifests itself as a change in shape either side of $150$~GeV. We note that the shape of the
$W +$~jets background is relatively stable when going from the exclusive to inclusive cuts. 
 
\begin{center} 
\begin{figure} 
\includegraphics[width=7cm]{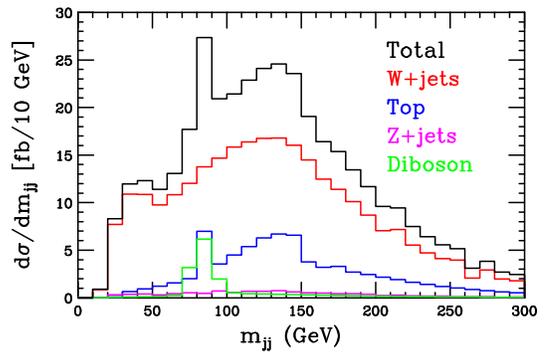}
\caption{NLO predictions for $m_{jj}$ using the ``inclusive'' CDF cuts (two or more jets), with an
increased jet threshold, $p_{T}^j > 40$~GeV. The labeling
is as in Fig.~\ref{fig:mjj30Ex}.} 
\label{fig:mjj40Inc}
\end{figure}
\end{center}

To further study the relative importance of the top background we present results for the cross sections and $m_{jj}$ distribution with an increased 
minimum jet-$p_{T}$ cut in Table~\ref{CrosssecInc} and Fig.~\ref{fig:mjj40Inc}. Here we use the same inclusive cuts but increase the minimum
jet $p_T$ to $40$~GeV. This results in a further enhancement of the top background and a clear peak in the total SM distribution
in the $100-150$~GeV region. 
The shape of the $W +$~jets background has altered significantly as a result of the increased cuts, also peaking around 
140 GeV. Therefore, although these cuts may be useful to constrain the normalisation of the SM backgrounds,
they may lead to a reduction in significance of any potential signal in this region.

In ref.~\cite{Plehn:2011nx} the SM peak in the top distribution was suggested as an explanation of the CDF excess. 
We note that the peak in the top contribution at $m_W$ is
correlated with the feature at $140$~GeV since they both arise solely from semi leptonic $t\overline{t}$ events.
Since, under the inclusive cuts, a large fraction of the total $W$ peak is comprised of top events, one would expect good control
of the top normalisation if the $W$ peak is well-described. One way that the top background normalisation could be constrained
would be to observe and measure the $W$ resonance with inclusive cuts, particular with  a higher jet threshold such as $p_{T}^j > 40$~GeV. 
\renewcommand{\baselinestretch}{1.6}
 \begin{table}[]
 \begin{center}
 \begin{tabular}{|c|c|c|c|}
 \hline
 Process &$\sigma^{LO} $~[fb] &  $\sigma^{NLO} $~[fb] & Ratio (NLO/LO) \\
 \hline
 $W+2j$ 						 & 4984(8)$^{+41\%}_{-27\%}$    & 	5132(24)$_{-7\%}^{+5\%}$ & 1.03	\\
 \hline 
 $Z+2j$  						&   213(1)$^{+42\%}_{-27\%}$   & 216(1)$_{-8\%}^{+4\%}$  	 & 1.01 	\\
 \hline 
 $WW(\rightarrow q\overline{q}) $  &  142.2(4)$^{+8\%}_{-7\%}$  &  221.9(4)$^{+6\%}_{-4\%}$	 & 1.56	\\
 \hline 
 $WZ(\rightarrow q\overline{q}) $     &  27.24(8)$^{+9\%}_{-8\%} $   & 41.8(1)$^{+5\%}_{-5\%}	$ &1.53	\\
 \hline
 $ZW(\rightarrow q\overline{q}) $  &     5.11(2)$_{-9\%}^{+10\%}$ & 8.02(7)$^{+6\%}_{-4\%}$ & 1.57	\\
  \hline 
 $t\overline{t}$ (fully-$\ell$)  	&  48.5(4)$^{+46\%}_{-28\%}$	& 59.44(8)$^{+0\%}_{-8\%}$	  &  1.23	\\
 \hline 
 $t\overline{t}$ (semi-$\ell$)  	&   99.1(7)$^{+47\%}_{-27\%}$	  &   91.7(8)$^{+0\%}_{-11\%}$  & 0.93 \\
 \hline
 Single $t$ (s)				  &  25.92(4)$^{+10\%}_{-8\%} $  & 35.6(4)$_{-3\%}^{+3\%}$	 & 	1.37  \\
 \hline
 Single $t$ (t)				  &   61.0(1)$^{0\%}_{-2\%}$  & 49.4(1)$^{-1\%}_{+4\%}$	 & 	0.81 \\
 \hline
 \end{tabular}
  \renewcommand{\baselinestretch}{1.0}
 \caption{LO and NLO predictions for cross sections using the CDF cuts (exactly two jets). The 
 percentage theoretical uncertainty is estimated by varying the scale choice in the calculation by
 a factor of two about the central value of $2m_W$. Statistical uncertainty resulting from Monte Carlo
 integration is shown in parentheses as the error on the final digit.}
 \label{CrosssecEx}
 \end{center}
 \end{table}
  \renewcommand{\baselinestretch}{1.0}

\renewcommand{\baselinestretch}{1.6}
 \begin{table}[]
 \begin{center}
 \begin{tabular}{|c|c|c|c|}
 \hline
 Process &$\sigma^{LO} $~[fb] &  $\sigma^{NLO} $~[fb] & Ratio (NLO/LO) \\
 \hline
 $W+2j$ 						&  4984(8)$^{+41\%}_{-27\%}$    & 	 5704(24)$^{+9\%}_{-13\%}$ & 1.14	\\
 \hline 
 $Z+2j$  						   & 213(1)$^{+42\%}_{-27\%}$   & 236(2)$^{+8\%}_{-12\%}$	 & 1.11	\\
 \hline 
 $WW(\rightarrow q\overline{q}) $    & 142.2(4)$^{+8\%}_{-7\%}$  &  252.3(8)$^{+8\%}_{-6\%}$ & 1.75	\\
 \hline 
 $WZ(\rightarrow q\overline{q}) $        &  27.24(8)$^{+9\%}_{-8\%} $   & 47.76(12)$^{+8\%}_{-7\%}$	 &1.75	\\
 \hline
 $ZW(\rightarrow q\overline{q}) $    &  5.11(2)$_{-9\%}^{+10\%}$   & 	9.02(2)$_{-7\%}^{+9\%}$ & 	1.77\\
  \hline 
 $t\overline{t}$ (fully-$\ell$)  	   & 48.5(4)$^{+46\%}_{-28\%}$	& 67.1(1)$^{+4\%}_{-11\%}$	  &  1.38	\\
 \hline 
 $t\overline{t}$ (semi-$\ell$)  	  &  686.9(1)$^{+45\%}_{-29\%}$	  &    674.2(1)$^{+3\%}_{-11\%}$  &0.98 	\\
 \hline
 Single $t$ (s)				    & 25.92(4)$^{+10\%}_{-8\%} $  & 41.68(4)	$^{+7\%}_{-5\%}$ & 1.61	  \\
 \hline
 Single $t$ (t)				    &  61.0(1)$^{0\%}_{-2\%}$  &  59.8(1)$^{+1\%}_{-0\%}$ &0.98 	\\
 \hline
 \end{tabular}
  \renewcommand{\baselinestretch}{1.0}
 \caption{
 LO and NLO predictions for cross sections using the ``inclusive'' CDF cuts (two or more jets). 
 Uncertainties are calculated and indicated in the same fashion as for Table~\ref{CrosssecEx}.}
 \label{Crosssec30Inc}
 \end{center}
 \end{table}
  \renewcommand{\baselinestretch}{1.0}
\renewcommand{\baselinestretch}{1.6}
 \begin{table}[]
 \begin{center}
 \begin{tabular}{|c|c|c|c|}
 \hline
 Process &$\sigma^{LO} $~[fb] &  $\sigma^{NLO} $~[fb] & Ratio (NLO/LO) \\
\hline
 $W+2j$ 						  & 2568(4)   & 2784(16)  & 1.08	\\
 \hline 
 $Z+2j$  						   &   104.6(8) &  112(1)	 & 1.07	\\
 \hline 
 $WW(\rightarrow q\overline{q}) $    &  66.6(1)&  131.4(4)	 & 1.98	\\
 \hline 
 $WZ(\rightarrow q\overline{q}) $     &  14.56(4)& 27.96(8) &  1.92	\\
 \hline
 $ZW(\rightarrow q\overline{q}) $    &    2.28(1)  & 4.56(2)	 & 2.00	\\
  \hline 
 $t\overline{t}$ (fully-$\ell$)  	 & 38.2(8) 	& 53.92(8)  &  1.41	\\
 \hline 
 $t\overline{t}$ (semi-$\ell$)  	  &  	655.0(7) & 642.2(7)     & 0.98	\\
 \hline
 Single $t$ (s)				    & 19.44(4) &  30.96(4)	 & 	1.59  \\
 \hline
 Single $t$ (t)				  &43.36(8) & 42.20(8)	 & 0.97	\\
 \hline
 \end{tabular}
  \renewcommand{\baselinestretch}{1.0}
 \caption{LO and NLO cross sections for  the $p_{T}^{j} > 40$ inclusive final state. Scales are set at $\mu_F=\mu_R=2m_W$.}
 \label{CrosssecInc}
 \end{center}
 \end{table}
In order to compare our NLO calculations with predictions obtained using tools that are commonly-used experimentally,
we present results obtained using a combination of ALPGEN~\cite{Mangano:2002ea} and Pythia~\cite{Sjostrand:2006za}. 
Although only accurate to leading order in the total cross section, the parton shower provides a
more realistic environment in which to perform a jet veto.
Here we will concentrate on the two most important backgrounds,  those from $W+ $ jets and top.
 For the $W+$~jets background we use a matched set of events, while the top backgrounds simply apply the
parton shower to a single set of tree-level matrix elements.
For the parton shower, particles are formed into jets using the midpoint cone algorithm ($R=0.4$) via 
FastJet~\cite{Cacciari:2005hq} and we use the CTEQ6L PDF set~\cite{Pumplin:2002vw}.

\begin{center} 
\begin{figure} 
\renewcommand{\baselinestretch}{1.0}
\includegraphics[width=7cm]{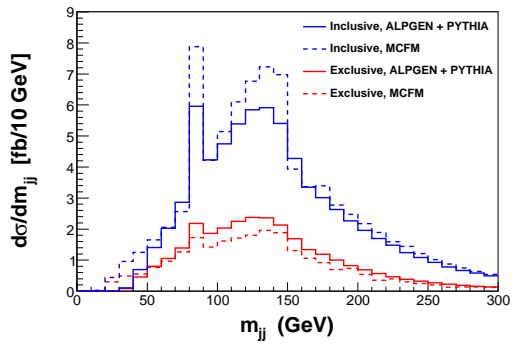}
\caption{Comparisons of the top backgrounds using MCFM (dashed) and ALPGEN + Pythia (solid) for the exclusive (red) and ``inclusive''  (blue) cuts.} 
\label{fig:Alp30EX}
\end{figure}
\end{center}

We first compare the top distribution under the exclusive and inclusive cuts (with $p_{T}^j >$ 30 GeV) in Fig.~\ref{fig:Alp30EX}.
To best compare the shapes we have adjusted the distribution obtained from the parton shower such that the $W$ peak
is aligned with the parton-level calculation, thus partially correcting for fragmentation and hadronisation effects.
As expected from the small corrections to the top processes at NLO, the normalisation of this background is in approximate agreement
between the two approaches.  However the parton shower gives rise to a somewhat different shape, particularly in the inclusive case
where the peak around $140$~GeV is broadened.

 \begin{center} 
\begin{figure} 
\includegraphics[width=7cm]{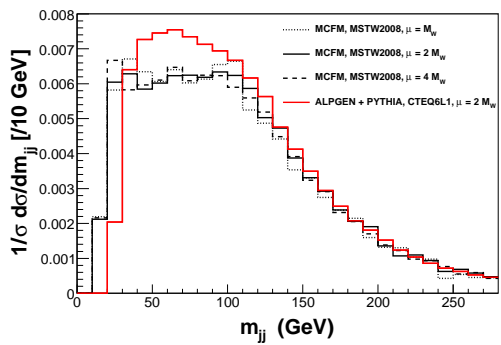}
\renewcommand{\baselinestretch}{1.0}
\caption{Comparisons of the $W+2j$ backgrounds using MCFM and ALPGEN + Pythia for the  ``inclusive''  cuts. At NLO the dependence of the shape upon the scale choice is shown by the dotted and dashed curves. } 
\label{fig:W_scal}
\end{figure}
\end{center}
\renewcommand{\baselinestretch}{1.0}
The other crucial background process is $W+$~jets, for which we compare the results from MCFM and the parton shower in Fig.~\ref{fig:W_scal}.  
We present the NLO and showered results normalised to their own cross sections so that we can compare the relative shapes. 
We observe that the change in the shape of the NLO calculation as the scale is varied is small.
 The prediction from the parton shower has a similar shape as
the parton-level results in the tail but differences appear at lower $m_{jj}$. However this is precisely the region in which
we would expect the fixed order calculation to begin to break down and the parton shower to be more reliable.

\section{Conclusions}
\label{sec:conc} 

We have presented NLO predictions for cross sections and dijet invariant mass distributions for one lepton,
missing $E_T$ and two jets at the Tevatron.  We have used a variety of cuts, including those  used by the CDF
collaboration who have recently reported an excess in this distribution around 150 GeV. 
By calculating the distribution of the invariant mass of the dijets at NLO we have ruled out large NLO $K$-factors as 
a possible source of the excess within the context of the SM. At NLO the cross sections have only a moderate dependence on the
renormalisation and factorisation scales of QCD, indicating that our results could be used to constrain the overall normalisation
of these backgrounds. 

The SM predicts a parton-level edge in the top background around $150$ GeV, an edge that is softened into a
broader peak by the parton shower.
Detector effects, that we have not considered here, will certainly modify this feature further. In order
to gain better control over the shape of this background we would advocate the use of the more inclusive
cuts for which the top background is much larger and thus more easily constrained. It becomes even more significant
for cuts demanding harder jets. For instance a $p_{T}^j > 40$~GeV cut yields a top cross section in the region
of the broad peak only a factor of $2.5$ lower than the $W+$~jets contribution.
Further information on this background could be gleaned by investigating the dijet mass distribution 
for the case of $b$-tagged (or anti-tagged) jets. In particular, the dominant source of two anti-$b$-tagged
jets is from the hadronic decay of the $W$. The invariant mass of two anti-$b$-tagged jets should therefore peak
sharply around $m_W$, with no significant peak in the $100$--$150$ GeV region.

\section{Acknowledgments}

We thank Viviana Cavaliere, Keith Ellis, Walter Giele, Roni Harnik, Joey Huston, Graham Kribs, Fabio Maltoni and Zack Sullivan and Jan Winter for useful discussions.
Fermilab is operated by Fermi Research Alliance, LLC under Contract No. DE-AC02-07CH11359 with the United States Department of Energy.

\bibliography{CDFdijet}

\end{document}